\documentclass{ws-jai}
\usepackage[flushleft]{threeparttable}
\usepackage{enumerate}
\usepackage{color}
\newcommand{\mm}{~$\mu\text{m}$}
\newcommand{\ap}{$\sim$}

\begin{document}

\catchline{}{}{}{}{} % Publisher's Area please ignore

\markboth{A.~M.~Glauser}{EChO Instrument}

\title{Characterizing Exoplanets in the Visible and Infrared: \\A Spectrometer Concept for the EChO Space Mission}

\author{A.~M.~Glauser$^{1*}$, R.~van~Boekel$^1$, O.~Krause$^1$, Th.~Henning$^{1}$, B.~Benneke$^{2}$, J.~Bouwman$^{1}$, P.~E.~Cubillos$^{3,1}$, I.~J.~M.~Crossfield$^{1}$, \"O.~H.~Detre$^{1}$, M.~Ebert$^{1}$, U.~Gr\"ozinger$^{1}$, M.~G\"udel$^{4}$,  J.~Harrington$^{3,1}$, K.~Justtanont$^{5}$, U.~Klaas$^{1}$, R.~Lenzen$^{1}$, N.~Madhusudhan$^{6}$, M.~R.~Meyer$^{7}$, C.~Mordasini$^{1}$, F.~M\"uller$^{1}$, R.~Ottensamer$^{4}$, J.-Y.~Plesseria$^{8}$, S.~P.~Quanz$^{7}$, A.~Reiners$^{9}$, E.~Renotte$^{8}$, R.-R.~Rohloff$^{1}$, S. Scheithauer$^{1}$, H.~M.~Schmid$^{7}$, J.-R.~Schrader$^{10}$, U.~Seemann$^{9}$, D.~Stam$^{10}$, B.~Vandenbussche$^{11}$, and U.~Wehmeier$^{7}$}

\address{
$^1$Max Planck Institute for Astronomy, K\"onigstuhl 17, 69117 Heidelberg, Germany\\
$^2$Massachusetts Institute of Technology, 77 Massachusetts Avenue, Cambridge, MA 02139, USA\\
$^3$Planetary Sciences Group, Department of Physics, University of Central Florida, Orlando, FL 32816, USA\\
$^4$University of Vienna, Department of Astrophysics, T\"urkenschanzstr. 17, 1180 Wien, Austria\\
$^5$Onsala Space Observatory, 439 92 Onsala, Sweden\\
$^6$Yale University, 260 Whitney Avenue, New Haven, CT 06511, USA\\
$^7$ETH Zurich, Institute for Astronomy, Wolfgang-Pauli-Str. 27, 8093 Zurich, Switzerland\\
$^8$Centre Spatial de Li\`ege, Parc Scientifique du Sart Tilman, Avenue du Pr\'e-Aily, 4031 Angleur-Li\`ege, Belgium\\
$^9$Institut f\"ur Astrophysik G\"ottingen, Friedrich-Hund-Platz 1, 37077 G\"ottingen, Germany\\
$^{10}$SRON, Sorbonnelaan 2, 3584 CA Utrecht, The Netherlands\\
$^{11}$Instituut voor Sterrenkunde, Katholieke Universiteit Leuven, Celestijnenlaan 200D, 3001 Leuven, Belgium\\
$^*$glauser@mpia.de
}

\maketitle

\begin{history}
\received{2013 March 22};
\revised{2013 May 13};
\accepted{2013 May 13}
\end{history}

\begin{abstract}
Transit-spectroscopy of exoplanets is one of the key observational techniques to characterize the extrasolar planet and its atmosphere. The observational challenges of these measurements require dedicated instrumentation and only the space environment allows an undisturbed access to earth-like atmospheric features such as water or carbon-dioxide. Therefore, several exoplanet-specific space missions are currently being studied. One of them is EChO, the Exoplanet Characterization Observatory, which is part of ESA's Cosmic Vision 2015-2025 program, and which is one of four candidates for the M3 launch slot in 2024. 

In this paper we present the results of our assessment study of the EChO spectrometer, the only science instrument onboard this spacecraft. The instrument is a multi-channel all-reflective dispersive spectrometer, covering the wavelength range from 400~nm to 16{\mm} simultaneously with a moderately low spectral resolution. We illustrate how the key technical challenge of the EChO mission - the high photometric stability - influences the choice of spectrometer concept and drives fundamentally the instrument design. First performance evaluations underline the fitness of the elaborated design solution for the needs of the EChO mission.
\end{abstract}


\section{Introduction}
One of the most exciting developments in modern astronomy was the detection of more than 850 extrasolar planets in the last two decades. And with the first detections of transiting exoplanets by \citet{Charbonneau:2000fh} and \citet{Henry:2000bc}, this new field became even more fascinating as it opened the possibility to characterize the exoplanet and its atmosphere: From the light curves observed during transit events, the planet-to-star size ratio can be accurately determined. In combination with radial velocity measurements and estimates of the stellar mass and radius, this yields direct measurements of the planetary size and mass, with accuracies that are in practice limited only by the uncertainties of the corresponding stellar parameters. Thus, the planet's mean density can be derived, placing constraints on the bulk composition and interior structure. Comparison to theoretical planet structure models provide estimates of the core mass and of the gaseous envelope, even though solutions are not unique.

The natural next step is to characterize directly the planet atmosphere, and transiting planets provide unique opportunities to do so. During a \emph{primary eclipse} event, also called a ``transit'', the planet passes in front of the star and the absorption signature of the planet atmosphere can be measured in transmission \citep[e.g.,][]{Charbonneau:2002er}. During a \emph{secondary eclipse} event, also called ``occultation'', the planet passes behind the star and the obscured planet radiation (emitted or reflected) can be measured \citep[e.g.,][]{Charbonneau:2005be,Deming:2005fg}. In both cases, the prime observable is the eclipse depth, i.e., the fraction by which the total system flux is reduced during an eclipse event, measured in one or more photometric bands or by a spectrograph. Spectroscopic observations provide the most detailed characterization of the composition and temperature structure. A third method is to measure the flux variation of the star-exoplanet system during a half or full orbital period. This phase-resolved information can be used to determine 1D or 2D maps of the planet brightness temperature \cite{Harrington:2006,Knutson:2007}, providing an even deeper insight in the physics and structure of the exoplanet and its atmosphere.
 
In all three cases, the signal of the planetary atmosphere is obtained by accurate temporal monitoring of the combined light since the planet and the star are not spatially resolved. The prime challenge in such observations is to overcome the huge contrast between the signals from the planetary atmosphere and the host star. This depends on several stellar and planetary parameters, but is only of the order of a few millimagnitudes in the most favorable cases, and often much less. Hence, extremely good (spectro-) photometric stability is required for such measurements and very robust instrumentation is needed. The signal-to-noise ratio (S/N), at which a spectrum of a planet atmosphere can be extracted, is ultimately limited by the finite number of photons that can be recorded during one or more eclipse events, assuming stellar variabilities to be low compared to the star brightness. However, systematic effects arising in the instrument or, in the case of ground-based observations, in the Earth's atmosphere, prevent photon noise limited S/N to be reached. On a more fundamental level, many wavelength ranges in the infrared are not accessible from the ground because the Earth's atmosphere is not transparent there.

A facility with optimized performance for exoplanet characterization must therefore be placed into space. Existing facilities such as the Hubble Space Telescope (HST) or the Spitzer Space Telescope were already very successful for the initial transit observations. However, the performance for transit spectroscopy with these observatories is limited, as they have not been built primarily for this science case: The past and present instruments of HST with spectrometric capabilities provide good coverage from UV to near-infrared wavelengths. However, they are not optimized for the longest wavelengths as HST has a warm telescope and is therefore background dominated for wavelengths greater than \ap 1.6\mm. {\sc Spitzer}, on the other hand provided a cryogenically cooled telescope with spectroscopic capabilities from 5{\mm} onwards with the InfraRed Spectrograph (IRS). However, the choice of slit-width in combination with the pointing accuracy caused varying slit-losses \cite{Swain:2008}. Together with drift-effects of the Blocked-Impurity-Band (BIB) detectors \cite{Deming:2006, Agol:2010}, the data reduction of long time series requires sophisticated de-correlation methods to remove systematic variations of the photometric signal. \citet{Gibson:2011} showed how the determination of systematic noises is crucial to  prevent misinterpretations of the finally reduced spectra. By design, neither HST nor {\sc Spitzer} provide simultaneous coverage of a large wavelength range. This means for transit spectroscopy that either only a part of the wavelength range is accessible or observations of multiple transits are required. In combination, the important wavelength range between \ap2.0{\mm} and 5.0{\mm} is presently not accessible by space telescopes nor fully observable by ground-based observatories due to absorption by various molecular species in the Earth's atmosphere (mainly H$_2$O, CH$_4$, CO$_2$). Even future missions such as the James Webb Space Telescope (JWST) will not provide simultaneous spectral coverage in the accessible wavelength range (0.6~-~14/28{\mm}) and are therefore susceptible to long term instrumental and stellar variability.

The Exoplanet Characterization Observatory (EChO) is a mission concept specifically designed to simultaneously observe the visible to mid-IR spectrum of transiting exoplanets \cite{Tinetti:2012hz}. EChO has been proposed as an M3 mission candidate for the Cosmic Vision 2015-2025 program of the European Space Agency (ESA). EChO's spectral coverage (400~nm to 16{\mm}) and moderate resolution is sufficient to simultaneously capture all spectral observables necessary to uniquely constrain the atmosphere of transiting exoplanets \cite{Benneke:2012}. The observatory is optimized for photon-noise limited transit-spectroscopy for targets brighter than K\ap9-10~mag. We present in Sect.~\ref{sec:mission_overview} the key figures of the EChO mission and the high level requirements. A first design study was conducted with the ESA Concurrent Design Facility (CDF) \cite{Puig:2011bz}. In the following Phase 0/A, the mission study continued with two parallel industrial studies of the spacecraft system and with two nationally funded instrument studies \cite{Puig:2012dc}. In this paper we present the results of one of these instrument studies, as introduced by \citet{Krause:2012el}. This instrument study is based on a detailed review of generic spectrometer concepts for the specific application of transit spectroscopy. In particular, the high constraints on instrumental stability require a careful assessment of instrumental concepts, design and technologies. From the photometric stability requirement we derive in Sect.~\ref{sec:trade-off} an optimized instrument concept and key instrumental parameters. The corresponding design solution is thereafter presented in Sect.~\ref{sec:design_description}. We provide first estimates of the expected performance in Sect.~\ref{sec:performance}. A summary and brief outlook are described in Sect.~\ref{sec:outlook}.

%===================================================================================================

\section{Mission Overview}\label{sec:mission_overview}
The EChO mission will provide low-resolution spectrometric capabilities for a broad wavelength range from visible to mid-infrared simultaneously. An overview of the key mission requirements of EChO is shown in Table~\ref{tab:key_figures_ECHO} (for a complete set of requirements, see the EChO Mission Requirement Document \cite{ECHOMRD}).  
\begin{table}[h]
\caption{Key figures of the EChO mission (status Sept. 2012)}
\label{tab:key_figures_ECHO}
\begin{center}
\begin{tabular}{@{}ll@{}}
\toprule
Parameter & Value\\
 \colrule
Wavelength coverage: 		& 550~nm - 11{\mm} (requirement)\\
						& 400~nm - 16{\mm} (goal)\\
Spectral Resolving Power:	& 300 ($\lambda\le 5\mu$m)\\
						& \hphantom{0}30 ($\lambda> 5\mu$m) \\				
Sensitivity limit:				& Brightest star limit (stellar type):\\
						& \hphantom{0}K0V, $K_\textrm{mag}=4.0$ (requirement)\\
						& \hphantom{0}F9V, $K_\textrm{mag}=2.9$ (goal)\\	
						& Faintest star limit (goal: $K_\textrm{mag} +1$):\\
						& \hphantom{0}M5V, $K_\textrm{mag}=8.8$ ($\lambda<3\mu\textrm{m}$)\\
						& \hphantom{0}G0V, $K_\textrm{mag}=9.0$ ($3\mu\textrm{m}\le\lambda\le8\mu\textrm{m}$)\\		
						& \hphantom{0}G0V, $K_\textrm{mag}=8.8$ ($8\mu\textrm{m}<\lambda$)\\										
System noise limit:			& $X\le2.0~(\lambda<1\mu\textrm{m}$)\\
(see Eq.~(\ref{eq:X}))			& $X\le0.3$ (goal: $X\le0.1$) $(\lambda\ge1\mu\textrm{m}$)\\					
Time resolution:			& 90~seconds (30~seconds goal)\\						
Typical observation length:	& 10~hours\\
Telescope:				& 3-mirror, Korsch, afocal, off-axis\\	
						& Diameter M1: 1.2~m\\			
						& Effective area: 1.131~m$^2$\\
						& Entrance / exit pupil size: 1286.5~mm / 36.6~mm\\
						& Effective focal length: 10568.3~mm\\
						& (with focusing lens with 300~mm of focal length).\\
Telescope temperature:				& 45~K / 55~K, depending on cooling concept\\
Launcher + Orbit:			& Soyuz, L2\\										
Launch Date:				& 2024 (or 2022, depending on L-class programatic)\\
\botrule
\end{tabular}
\end{center}
\end{table}
This space-observatory is dedicated for transit spectroscopy of a wide range of stellar and planetary targets. The mission sample includes hot to temperate Jupiters, Neptuns and even hot to warm super Earths. The broad spectral range provides access to various molecular features such as methane, water, ammonia, carbon di- and monoxide, oxygen, ozone, and many others. The science case and mission sample selection are summarized in the Science Requirement Document \cite{ECHOSciRD}.

The observatory shall allow differential spectroscopy of transiting exoplanets with a relative photometric accuracy that is fundamentally limited only by the observed astronomical object (photon noise of target and zodiacal background). This guarantees the maximum sensitivity for star-to-planet contrast ratios with a given telescope effective area. The consequence of such a mission goal is that the observatory has to provide superb conditions for photometric stability over a long time duration of hours (for transit measurements) to days (for phase-resolved observations). In Sect.~\ref{sec:trade-off} we explore further the consequences of this requirement.

The EChO telescope baseline consists of an off-axis, afocal three-mirror Korsch telescope with effective area of 1.131~m$^2$. The telescope will be cooled down passively to 45~K or 55~K, depending on the instrumental cooling concept. To provide a thermal environment with sufficient stable performance and to maximize the sky visibility, the orbit of EChO will be located around the second Lagrangian point (L2).

EChO will host only one science instrument and will be operated presumably in only one observation mode: Long-time series spectroscopy of point-sources. This simplifies the observatory architecture significantly as, e.g., no cryo-mechanisms are needed.

%===================================================================================================

\section{Photometric Stability - A Key Design Driver}\label{sec:trade-off}
As introduced in Sect.~\ref{sec:mission_overview}, the achievable S/N shall be limited only by the photon noise of the stellar object itself (or the zodiacal background in case of faint objects and long wavelengths). This  requirement is the biggest challenge of the EChO mission, as it demands for an observatory with extremely high photometric stability and sophisticated data analysis methods. Since transit-spectroscopy is conducted by comparing the spectrophotometric signal during and after/before the eclipse, the relative photometric response function has to be quasi-constant over the observation length, typically 10~hours. As no instrument will allow such perfect conditions, it is important to quantify instrumental effects that will lead to modulations of the photometric response and increase the noise floor of the resulting differential spectrum, consequently. We illustrate in this section how the requirement for photometric stability drives the definition of critical instrumental parameters and finally determines the optimum spectrometer concept. 

\subsection{Quantifying the photometric stability}\label{sec:xfact}

First, we derive the quantitative requirement for the acceptable instrumental noise: As introduced in the Mission Requirement Document \cite{ECHOMRD} the total instrumental noise is expressed by the quantity $X$, which is a relative quantity compared to the photon noise. The total noise $\sigma_{\textrm{Total}}$ can be written as 
\begin{equation}\label{eq:X}
\sigma_{\textrm{Total}}=\sqrt{N\times(1+X)},
\end{equation}
while $N$ is the number of collected electrons from the target and any non-instrumental background (such as the zodiacal background) in a given time interval. For $X=0$, the S/N is determined only by the photon noise (of non-instrumental sources). For $X=1$, the observation is not photon noise dominated anymore. $X$ is the sum of all possible instrumental relative noise terms $\sigma_i$ with $X=\sum\frac{\sigma_i^2}{N}$, caused by effects such as detector dark current, thermal emission of the telescope, drifts in throughput, and so forth. As indicated in Table~\ref{tab:key_figures_ECHO}, $X$ is required to be smaller than 0.3 (goal 0.1) for wavelengths larger than 1{\mm} and 2.0 for shorter wavelengths. A limit of $X\le0.3$ comprises a good margin for photon noise limited performance but also accounts for technical feasibility. This is different for the visible channel as for the cooler stellar targets, their black-body emission is in the Wien regime where a fast drop in intensity is expected. Further,  50\% of the light will not be available as it will be used for the Fine Guidance System (see Sect.~\ref{sec:optchain}). Therefore, photon noise limited performance is very difficult to achieve in this wavelength band and the relaxed requirement $X\le2.0$ takes this into account, accordingly.

$X$ has to be quantified and budgeted taking into account potential correction methods and residual noise of calibration techniques. We used our Dynamic Performance Calculator to model the systematic and fundamental noises and their propagation in the overall system performance (see Sect.~\ref{sec:noisebudget}). To do this, the following steps are required: 1. Identification of systematic and fundamental noise sources, 2. Study of how these noise sources propagate in the resulting noise performance (noise type), and 3. Quantification of acceptable limits iteratively using the Dynamic Performance Calculator.

First, we address the three different types of noise sources and how they translate into the formalism of the $X$-term. We use the formalism introduced in Eq.(\ref{eq:X}) by describing noise contributions $\sigma_i$ as $X_i=\sigma_i^2/N$, with $N$ as the total collected signal electrons per spectral resolution element of the astronomical source (star + zodiacal background):
\begin{enumerate}[    A.]

\item Noise sources, which modulate the incoming flux linearly and are independent of the integration time. They scale with $\sigma_i=a\times N_0$, with $N_0$ is detected electrons/second (for the signal of the target and zodiacal background only) and $a$ is the scale factor. The $X$-factor is then $X_i=a^2\times N_0/ T$ with $N=N_0\times T$, and $T$ is the observing time. As an example, high frequent throughput variations qualify as type A noise sources, if no systematic drift is present. Otherwise, an additional component of type B noise has to be considered. 

\item Noise sources which are caused by systematic uncertainties (e.g., residual error after de-trending systematic drifts in the raw time series). These errors lead to a limitation of the achievable signal-to-noise, hence $\sigma_i/N=c$ (constant) for long observation durations $T$, or $\sigma_i=c\times N_0\times T$. The $X$-factor scales with $T$: $X_i=c^2\times N_0\times T$.

\item Noise of constant background signals with Poisson characteristics (e.g. telescope background, dark current). They scale with the square root of the observing time with $\sigma_i=\sqrt{B_0\times T}$ ($B_0$ are the detected electrons/second of the background source) and the $X$-factor is $X_i=B_0/N_0$, constant with time.
\end{enumerate}

\noindent This list shows that the nature of the noise determines the time scale of the impact on the resulting performance: Photometric instabilities caused by, e.g., slit-losses (see Sect.~\ref{sec:slitloss}) might modulate the signal strongly within the time scale of the telescope pointing-jitter. However, if the modulation is purely random (noise type A), it will have less impact on the resulting S/N if the integration lasts much longer than the noise time scale. But, if the modulation also contains systematic errors even after de-trending the time series, e.g., caused by the limited accuracy of the pointing information, the achievable S/N is limited regardless of the integration time (noise type B). This means in the terminology of $X$ that systematic errors deploy their dominance with increasing integration times ($X$ scales with $T$ linearly). Consequently, the acceptable limit for systematics is very low: In the extreme case for a 10~hour integration of a bright target, the maximum number of photons in a single spectral bin can be up to \ap10$^{10}$. To be compliant with the $X=0.3$ requirement, a residual systematic error of only \ap 5~ppm is allowed, assuming no other error sources are present!

The contribution of photon noise from the telescope background (type C) to the total noise scales like the photon noise of the astronomical target with $\sqrt{T}$. Consequently, the relative contribution to the S/N is independent of the integration time. Therefore, these noise sources have to be suppressed sufficiently for all time scales and target fluxes. This is challenging for the long wavelength range of EChO as the temperature of the telescope contributes significantly to $X$ for the case of faint targets (see Sect.~\ref{sec:noisebudget}).

\subsection{Basic spectrometer concept}
With the tool of the $X$-factor in hand we explored systematically the optimum spectrometer design for EChO. First, we conducted a fundamental review of possible spectrometer concepts to ensure that the challenging requirement of photometric stability is addressed best. We identified the following spectrometer concepts as possible candidates for the EChO instrument:
\begin{enumerate}[ (1)]
\item Multi-channel dispersive spectrometer with all-reflective optics (see Sect.~\ref{sec:dd_overview}): \\
6 channels separated by dichroic mirrors, each containing a grating covering one octave of the spectrum, camera optics, and a dedicated detector. A similar concept was introduced by the CDF study \cite{Puig:2011bz} using a mixture of prisms, gratings and optionally grisms as dispersive elements and a mixture of mirrors and lenses for the camera optics. For the trade-off between the different concepts however, we updated the multi-channel dispersive spectrometer concept with a design based purely on reflective optics (with the exception of the dichroic mirrors). This update was necessary due to the high susceptibility of refractive elements (prisms, grisms and lenses) to thermal variations, a critical parameter when considering the extreme requirements of photometric stability (see discussion at the end of this section). 

\item Cross-dispersive spectrometer:\\
Based on a similar approach to the multi-channel dispersive spectrometer, the cross-dispersive concept allows the reduction from 6 to 3 channels. Such a concept allows a more compact design and the reduction of the number of required detectors. Consequently, the electrical power and with it the thermal dissipation in the cryogenic system can be reduced. However, the cross-dispersion is provided by a prism and a grism in series, both refractive elements.

\item Static Fourier-Transform-Spectrometer:\\
A fundamentally different category of spectrometers is the Fourier-Transform-Spectrometer (FTS). In general, an FTS provides a high design flexibility (e.g., the resolving power can be easily adjusted by changing the optical path difference). Compared to the classical spectrometer the detection is feasible in the pupil plane. This provides a very stable measurement with respect to pointing-jitter (see Sect.~\ref{sec:sampling}). Similar to the multi-channel concept, the light is split up first in spectral octaves by dichroic mirrors. Each channel is then equipped with a Michelson interferometer. The static FTS concept is based on using fixed mirrors while the interferogram is generated by tilt of one of the interferometer mirrors according to the required resolution. Consequently, the optical path difference is translated into a physical dimension of the combined beam. By imaging the beam onto a detector array, an interferogram can be recorded.

\item Dynamic Fourier-Transform-Spectrometer:\\
The driver for studying also a dynamic (scanning) FTS was the benefit for long wavelengths in combination with Mercury Cadmium Telluride (MCT)  detectors: To prevent the need for active cooling, MCT detectors with long cut-off wavelengths up to 16{\mm} were considered. However, these detectors show detector dark current levels so high that the sensitivity requirement for photon noise limited performance could not be met for dispersive spectrometers or a static FTS. With a dynamic FTS however, the light is collected in only a few detector pixels, different optical path differences are realized by moving one mirror with a mechanism and the interferogram is collected over the scanning time. There is a much higher signal on each pixel, therefore a small dark current plays a minor role and thus makes long-wavelength MCT detectors a practicable solution. However, the need for a mechanism (moving mirror) is a potential source for failures but has to be weighted against the need for a cryo-cooler.

\item Fibre-fed pupil spectrometer:\\
This concept was proposed by ESA but not studied in detail by our group. The advantage of a fibre-fed pupil spectrometer is similar to the static FTS: The spectrometric measurement is conducted in the pupil plane, making it insensitive to pointing variations. However, to feed the fibre with the stellar light, the fibre has to be placed in the focal plane first where similar problems related to pointing-jitter are expected. Further, due to the limited Field of View (FoV) of such a system, it is expected that slit-loss effects are much more significant compared with the wide field mask proposed for the multichannel dispersive spectrometer (see Sect.~\ref{sec:slitloss}). Therefore, preliminary calculations showed a much higher susceptibility to pointing-jitter and this option was rejected.
\end{enumerate}
For each of these concepts (except the fibre-fed pupil spectrometer) an optical design was elaborated and implemented in a preliminary opto-mechanical model. We analyzed the concepts in terms of photometric stability, radiometric performance, susceptibility to detector effects, optical performance, calibration aspects, complexity and system risks, mechanical, thermal, and electrical characteristics, and programatic aspects such as technology readiness and development effort. In particular, the photometric stability assessment was further differentiated by studying the susceptibility of the spectrometer concepts to pointing-jitter, thermal variations and micro-vibrations. It is beyond the scope of this paper to describe the trade-off, the analyses and the results in detail. However, we provide here a summary of the key findings.

The susceptibility to pointing-jitter is most prominent where field masks are required (due to slit losses) and where the spectrum is obtained in the image plane (due to detector pixel gain variations, see Sect.~\ref{sec:sampling}). Consequently, spectrometers that do not require field masks (static and dynamic FTS) or image the pupil onto the detector (static FTS) provide significantly more pointing-invariant measurements. However, for photon noise limited systems and quasi-continuum spectra, an FTS (both static and dynamic) has intrinsic disadvantages compared with a dispersive concept: With similar photon conversion efficiencies (transmission, detector quantum efficiency) and spectral resolving powers, the achievable S/N of a dispersive spectrometer scales with the square root of the resolving elements faster compared to an FTS. This is the result that the Fellget advantage of an FTS disappears when photon noise shall be the dominant noise factor. The photon noise of the full continuum contributes to the noise of each individual spectral resolving element \cite{Maillard1988FTS}. Consequently, the sensitivity requirement for EChO can never be achieved with an FTS and both the static and the dynamic FTS concepts were therefore dismissed. 

The susceptibility to thermal variations of the system is most prominent for systems using refractive optics. While thermo-elastic effects can be compensated by a uniform choice of material for pure reflective optics, the change of refractive index with temperature modulates the optical characteristics of lenses, prisms and grisms significantly. Performance analyses have shown that the position of the spectrum on the detector array moves significantly with temperature changes of a few mK only. Therefore, the cross-dispersive concept is less suited for high photometric stability and was dismissed. Further disadvantages supported the deselection, such as the limited FoV due to the compact alignment of the cross-dispersed orders on the detector array.

Consequently, the classical multi-channel dispersive spectrometer concept was selected as baseline for this study despite its susceptibility to pointing-jitter.

\subsection{The impact of pointing-jitter}
For a classical dispersive spectrometer, the telescope pointing-jitter modulates the spectro-photometric response of the instrument in different ways. In the following, we summarize the most prominent effects and how they are addressed by the final design solution.

For the presented study, preliminary information for the expectable pointing-jitter were provided by ESA: The relative pointing error is in the order of 20~mas (1-$\sigma$) up to 90~sec and the performance reproducibility error is in the order of 10~mas from 90~sec to 10~hours.

\subsubsection{Slit-losses}\label{sec:slitloss}
The choice of slit size is crucial for any spectrometer: If the slit width is too narrow, the flux varies with the relative position of the point spread function (PSF) compared to the slit position because of the truncation of the outer part of the PSF, which still contains a significant part of the energy. For transit-spectroscopy, these slit-losses contribute significantly to the photometric stability budget. With a high quality calibration between position offset and resulting throughput, it is possible to de-correlate pointing information with the temporal-modulated signal. This allows minimizing the systematic noise (type B) of the slit-losses. For this purpose, very accurate pointing information is needed on a time scale similar to the read-out frequency of the detectors. Therefore, we plan to measure for all channels the position of the 0$^\textrm{th}$~order of the dispersive gratings simultaneously with the spectrum on the same detector array. With an estimated accuracy of 2~mas for the position determination, the residual noise contribution of type A and B terms of the slit-losses can be sufficiently controlled, assuming a slit width larger than 13~arcsec for observations up to 16{\mm} (see Sect.~\ref{sec:noisebudget}). This slit width corresponds approximately to the 5$^\textrm{th}$ airy ring of the diffraction limited PSF at 16{\mm}. 

Such a wide slit induces wavelength shifts with respect to pointing offset perpendicular to the slit. With the direct position determination using the 0$^\textrm{th}$ order, these wavelength shifts can be corrected by on-ground data processing.

From a pure slit-loss perspective, it would be ideal to open the slit as much as possible (ideally slit-less). On the other hand, opening the slit too much leads to an increase of light contamination from thermal and zodiacal background. In principle, these contributions can be measured and subtracted from the raw spectrum: We plan an off-target background measurement by providing spatial coverage over a field length of 30~arcsec. With 13~arcsec slit width, the photon noise contribution of the background (type C noise with constant $X$) is compliant with the total error budget except for wavelengths larger than 13{\mm}, where it exceeds the permitted allocation for faint sources (see Sect.~\ref{sec:noisebudget}). Given that this wavelength range is only part of the mission goal and not requirement, this compromise is acceptable, though.

\subsubsection{Sampling}\label{sec:sampling}
Another example for signal modulations caused by pointing-jitter is the combination with inter- and intra-pixel variations in the effective quantum efficiency (QE) of the detector. For the same input flux, the number of collected electrons can vary, depending on the size and position of the PSF with respect to the pixels. This is caused by a variation of the QE on different scales; within a pixel and from pixel to pixel. For a flat illumination, pixel-to-pixel gain variations can be measured and corrected. For this purpose, we plan for an on-board flat field calibration source (see Sect.~\ref{sec:fcs}) to achieve a flat-field calibration better than 0.5\% on a spatial scale of \ap10 pixels.

\begin{figure}[h]
\begin{center}
\includegraphics[width=0.8\textwidth]{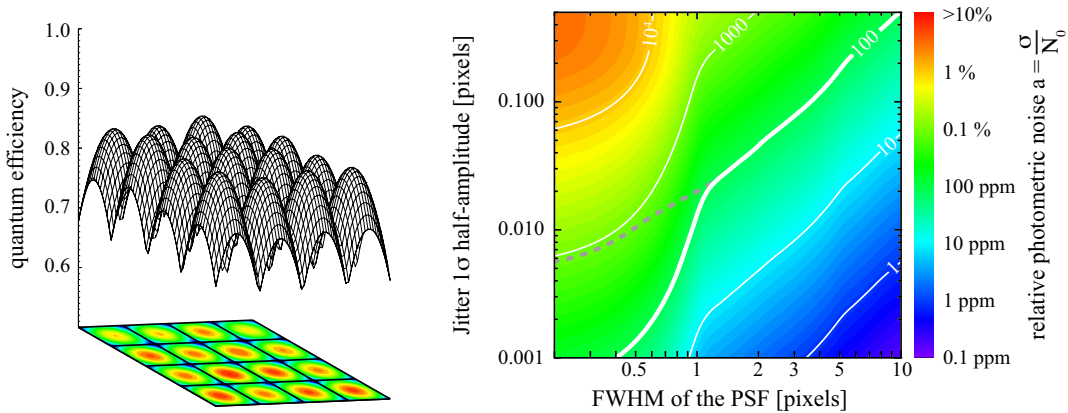} 
\end{center}
\caption{Illustration of the effects of PSF sampling on the photometric stability. Left panel: Modeled fine scale structure of the QE shown for 16 pixels. Right panel: The resulting noise in a simulated time series for a single integration as a function of PSF sampling and  pointing-jitter in units of pixels. The intra-pixel QE variation is modeled with 20\% (color axis and white contour) and 2\% (only shown for the 100~ppm line in gray-dashed).}
\label{fig:intra}
\end{figure}

The intra-pixel gain variation (as illustrated with arbitrary scale in the left panel of Fig.~\ref{fig:intra}) is an intrinsic effect of pixel array detectors where the QE is highest in the center of the pixel and falls off at the pixel edges \cite{Barron:2007}.
It is very demanding to determine the magnitude and topology of this characteristic. Furthermore, an in-orbit calibration of the fine scale structure of the gain distribution is unfeasible. Therefore, the instrument has to be designed to suppress this effect sufficiently by optimizing the sampling of the PSF with respect to plate scale and pointing-jitter. 

For this purpose, we studied the resulting photometric variation as a function of intra-pixel gain variation, pixel-to-pixel gain variation, pointing-jitter and PSF sampling. We assumed a pixel-to-pixel gain variation of 0.5\%, considering the residual calibration error of the flat field measurements. It is difficult to collect reliable information about the magnitude of the fine-scale QE topology. Therefore, we repeated the simulation for different magnitudes of the intra-pixel QE variation. The right panel of Fig.~\ref{fig:intra} shows the resulting relative photometric noise as a function of PSF sampling and pointing-jitter in units of pixels for an intra-pixel QE variation of 20\%. 

We aim to suppress the resulting photometric noise (type A) down to a level of $a<100$~ppm/integration ($a$ is the noise scale factor as introduced in Sect.~\ref{sec:xfact}) at which the impact on the resulting total system noise is not dominant: For long term drifts, de-correlation with the pointing determination is needed for which the long-term systematic effect (noise type B) can be reduced down to $c<3$~ppm ($c$ is introduced in Sect.~\ref{sec:xfact} as the constant ratio of noise to signal), given the direct position determination as described before. In Fig.~\ref{fig:intra}, the $a=100$~ppm contour line is shown for two scenarios of intra-pixel gain variations, the conservative value of 20\% (in solid white) and a very optimistic value of 2\% (dashed gray). This shall illustrate that for a sampling of the PSF by more than \ap2 pixels per full width of half maximum (FWHM), the resulting photometric variability is independent of the intra-pixel gain variation. Therefore, we set with some margin the minimum sampling width of the spatial PSF to 2.5 pixels.

This defines further the minimum pixel plate scale: Using the 100~ppm contour line of Fig.~\ref{fig:intra} and the pointing-jitter half-angle of 20 mas, a minimum plate scale of \ap 300~mas~/~pixel is required to sufficiently suppress this noise contribution. However, for a 1.2~m telescope, the diffraction limited PSF is too small to be sampled by 300~mas~$\times$~2.5~pixels for wavelengths smaller than \ap3.5{\mm}. For this reason, the PSF has to be artificially broadened to allow simultaneously a correct sampling rate and a sufficiently large pixel plate scale. We will show the hardware solution to that problem in Sect.~\ref{sec:optics}.

%===================================================================================================

\section{Instrument Design Description}\label{sec:design_description}

\subsection{Overview}\label{sec:dd_overview}
We present a design solution for the EChO science instrument which is compliant to achieve the mission goals. Table~\ref{tab:key_figures_instr} shows a summary of key parameters. In particular, the instrument concept is optimized for photometric stability, as introduced in Sect.~\ref{sec:trade-off}. 
\begin{table}[h]
\caption{Key figures of the EChO instrument}
\label{tab:key_figures_instr}
\begin{center}
\begin{tabular}{@{}l l@{}}
\toprule
Parameter&Performance\\
 \colrule
Wavelength coverage:	&	400~nm -- 16{\mm}, separated in 6 channels	\\
Spectrometer channels:	&	VIS (0.40 -- 0.80{\mm}), IR1 (0.70 -- 1.40{\mm}),\\
					&	IR2 (1.32 -- 2.64{\mm}), IR3 (2.58 -- 5.16{\mm}),\\
					&	IR4 (5.00 -- 9.00{\mm}), IR5 (8.60 -- 16.0{\mm})	\\
Spectral Resolution:		&	300 -- 600 for VIS, IR1, IR2, IR3\\
					&	100 -- 180 for IR4, IR5\\
FoV:					&	30 arcsec in spatial, 13 arcsec in spectral direction\\
Pixel plate scale:		&	385~mas~/~pixel for VIS, IR1, IR2, IR3\\
					&	802~mas~/~pixel for IR4, IR5\\
Detectors:				&	MCT (H2RG) for VIS, IR1, IR2, IR3			\\
					&	Si:As for IR4, IR5			\\				
Cooling:				&	55 K passive -- instrument base temperature										\\
					&	15 K / 7 K active with cryocoolers for IR4, IR5					\\
Total Mass:			&	97~kg for cold instrument unit 				\\
					&	67~kg for warm electronics and cooler compressors \\
Electrical Consumption:	&	34.2~W + 133~W for Cooler Electronics					\\					
\botrule
\end{tabular}
\end{center}
\end{table}%

\begin{figure}[h]
\begin{center}
\includegraphics[width=0.8\textwidth]{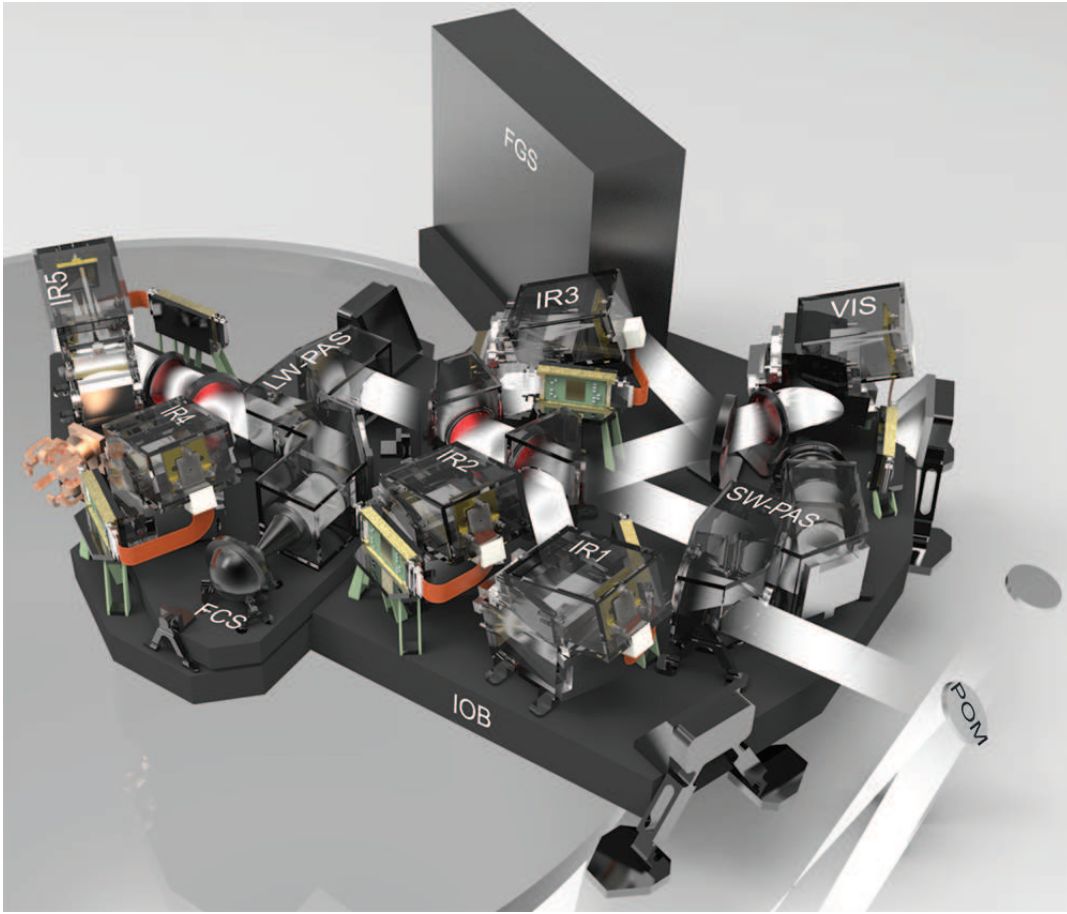} 
\end{center}
\caption{Overview of the EChO Science Instrument. Covers and thermal hardware are not shown.}
\label{fig:instr_overview}
\end{figure}

The baseline of the science instrument is a 6-channel dispersive spectrometer with all-reflective optics (one channel for the visible range, VIS, and 5 infrared channels, IR1-IR5). The instrument allows simultaneous spectral coverage from 0.4 to 16{\mm} with a spectral resolution of R~$\ge$~300 for wavelengths smaller than 5{\mm} and R~$\ge$~100 for wavelengths longer than 5{\mm}.  The FoV is the same for all channels, as only one field mask of 30 arcsec in the spatial and 13 arcsec in the spectral direction is used. The opto-mechanical implementation of the EChO science instrument is shown in Fig.~\ref{fig:instr_overview}. 

The detector types for the channels up to 5{\mm} are Mercury Cadmium Telluride (MCT) and require passive cooling to the instrument temperature at 55~K. To allow the spectral coverage up to 16{\mm}, Si:As detectors are foreseen for the channels IR4 and IR5, respectively. These detectors are operated at \ap 7~K and consequently, some parts of the instrument require active cooling. The Si:As detectors are sensitive up to 28{\mm}. Therefore, the instrument optics of channels IR4 and IR5 requires active cooling to about 15~K to suppress its thermal emission. 

The presented instrument design is based on components with very high technology readiness, hence no dedicated technology development program is required. However, additional development efforts are pushed in parallel to study European alternatives to the selected US detectors and front end electronics (see Sect.~\ref{sec:fpa}).

%---------------------
\subsection{Optical concept}\label{sec:optics}
\subsubsection{The optical chain}\label{sec:optchain}
\begin{figure}[h]
\begin{center}
\includegraphics[width=0.7\textwidth]{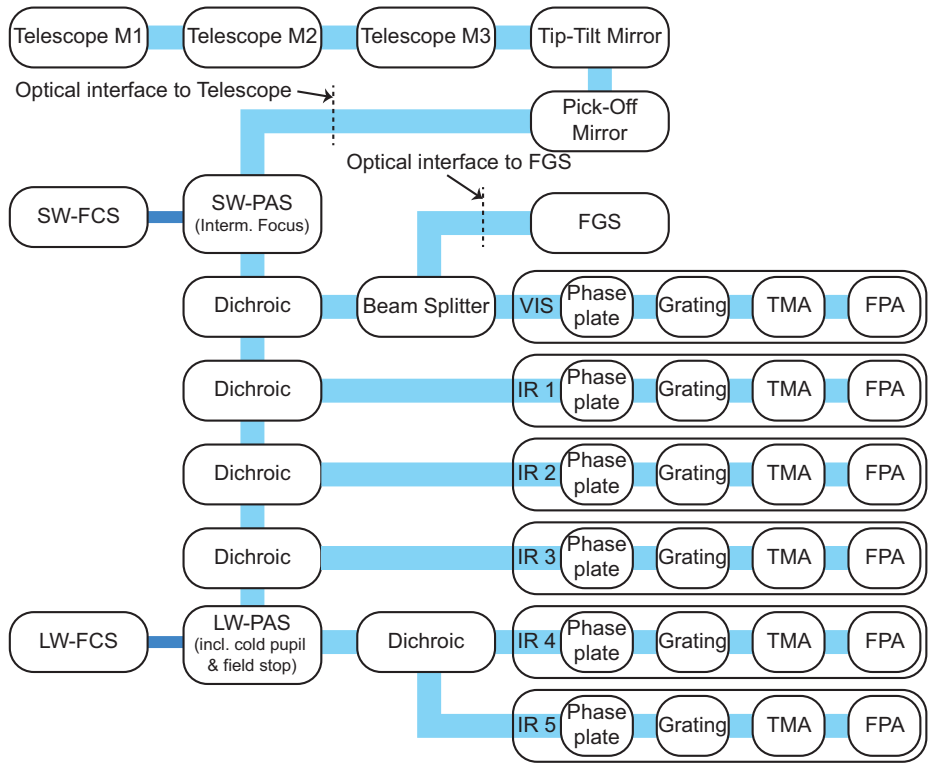} 
\end{center}
\caption{Optical architecture of the EChO Science Instrument}
\label{fig:opticalconcept}
\end{figure}

Fig.~\ref{fig:opticalconcept} illustrates the optical architecture for the EChO instrument.
After the last folding mirror of the a-focal telescope (Pick-Off Mirror, POM), the collimated beam passes through the short-wavelength Pupil Adapter System (SW-PAS), which comprises four off-axis mirror segments in symmetric arrangement. The first two mirrors feature biconic surfaces and deliver an intermediate focus with near-diffraction limited image quality. The subsequent two mirrors are standard conics and re-collimate the beam. The main purpose of the PAS is to provide the appropriate field mask and pupil stop, to reimage the telescope pupil onto the spectrometer gratings (for photometric stability), and to allow injecting a beam from an on-board calibration source used for flat-field measurements (FCS). With the current design of the telescope, the beam-size before and after the PAS remains unchanged (\ap40~mm in diameter) but the PAS features the robustness against future design changes, providing an invariant interface between the telescope and the instrument. As the spectrometer channels IR4 and IR5 require additional cooling and baffling against thermal radiation from the 55~K instrument environment, a second PAS (long-wavelength, or LW-PAS) is inserted, providing the interface between the 55~K and 15~K thermal environments. The optical design of the two PASs is very similar just taking into account the different positions of the pupils. 

Once the collimated beam passes the PAS, a series of dichroic mirrors is responsible for separating the different wavelength bands, guiding each to its dedicated spectrometer optics. For the VIS channel, an additional beam-splitter is installed to divide the beam equally to the spectrometer and the Fine Guiding System (FGS), which is a unit provided by the spacecraft but mounted on the instrument optical bench (IOB). 

For wavelengths shorter than \ap 3.5{\mm}, an artificial broadening of the PSF is needed to optimize the sampling with respect to photometric stability (see Sect.~\ref{sec:sampling}). To achieve this, a phase plate is inserted after the dichroic mirrors for each channel beam, acting as a low pass filter. The phase offsets of these plates follow a high frequency statistical distribution which can be tuned to suppress high-frequency spatial features in the resulting image and to broaden the PSF. The advantage of phase plates located close to the pupil plane compared with other PSF broadening approaches (such as de-focusing, birefringent filters, and scramblers, e.g., fibers) is their insensitivity against small pointing offsets. 

\begin{figure}[h]
\begin{center}
\includegraphics[width=0.4\textwidth]{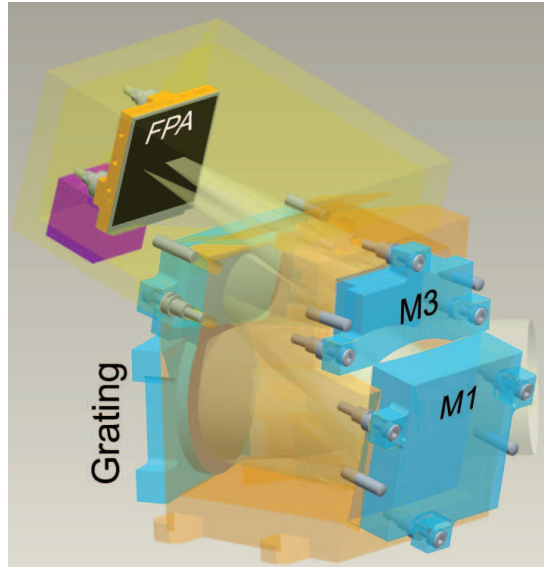} 
\end{center}
\caption{Three mirror anastigmats camera with grating and detector of channel IR1. The camera housing is indicated in half-transparency.}
\label{fig:tma}
\end{figure}

Each of the six spectrometer boxes consists of a grating and a three mirror anastigmats camera (TMA), followed by the focal plane array (FPA). The gratings are used in 1$^\textrm{st}$ order for spectroscopy and in 0$^\textrm{th}$ order for simultaneous imaging, providing the capability for accurate and simultaneous pointing monitoring (see Sect.~\ref{sec:slitloss}). Fig.~\ref{fig:tma} shows an opto-mechanical view of the spectrometer channel IR1. 
Although the cameras will be built close to diffraction limited performance, the resulting image quality will be dominated by the phase plates. Therefore, the camera design can be chosen identically for the channels VIS, IR1, IR2, and IR3 providing an f-ratio of 7.5. The channels for IR4 and IR5 are designed with an f-ratio of 5.0. This uniformity of the optics simplifies the manufacturing process and verification procedure substantially.

%---------------------
\subsubsection{Opto-mechanical implementation}
The main driver for the material selection is the very ambitious photometric stability requirement, which brings the focus onto thermo-elastic effects. Structures and reflective optics are all made out of the same material (Al6061) to avoid a mixture of different coefficients of thermal expansion (CTE). Such an approach is robust against thermal variations: For example, a reduction in the focal distance due to a temperature drift is compensated by the change in curvature radius of the mirrors. 

Diamond machining is the state of the art solution for the fabrication of reflective surfaces of this kind. With the advantage of the diamond tool due to its hardness, low wear, high chemical resistance, good thermal conductivity, and low friction, it is possible to create optical surfaces for infrared applications without the necessity of further post-polishing processes.

However, the VIS and IR1 spectrometer require a different coating technology for the reflective optics due to the performance limitation of diamond-turned aluminum at these wavelengths: An intermediate Ni-P layer provides the necessary stiffness and is base for the final gold, silver, or aluminum coating. The disadvantage of the Ni-P coating is that the CTE is different from Al6061, which causes bimetallic effects. Therefore a symmetric, thin Ni-P coating is necessary to minimize the bimetallic bending \cite{Rohloff:2010hb}.

To achieve sufficient alignment accuracy, the optics of the PASs and TMAs will be fabricated using a manufacturing technique that allows a ``snap-together'' integration. This technique is described by, e.g., \citet{Risse:2011ib} and is based on simultaneous machining of mirrors and its supporting structures. The achievable alignment precision is very high and repeatable, without the need for large post-manufacturing alignment efforts. 

%---------------------
\subsection{Focal plane array detectors}\label{sec:fpa}
Each of the spectral channels is equipped with a dedicated focal plane array (FPA) and front end electronics (FEE) unit. 
A survey of papers, investigations of test results and other accessible information confirmed that MCT detectors are the most appropriate detectors for nearly all of these wavelength bands. Due to their wide common use in ground-based and space-borne astronomy, the expectable performance of MCT detectors is well established.
The cut-off wavelength of MCT detectors can be tuned over a large wavelength range from the near infrared up to about 20{\mm} by adjusting the material composition.

In principle, MCT detectors could cover the required EChO wavelength range. However, due to the steeply increasing dark current towards longer cut-off wavelengths, Si:As Impurity Band Conduction detectors (IBC, Raytheon, see e.g., \citet{Ressler:2008kt}) or also so-called Blocked Impurity Band detectors (BIB, DRS) are considered for the long wavelength channels IR4 and IR5, in particular for covering the band up to 16{\mm}. The required low operating temperature for the Si:As IBC/BIB detectors of about 7 K demands the implementation of an active cooling system.

The FPA of the short-wavelength channels VIS, IR1, IR2, and IR3, respectively, consists of Teledyne's H2RG MCT detectors \cite{Blank:2012hb} operated at a temperature of 55 K, similar to the instrument optical bench. Their high technology readiness for space applications and outstanding performance make them most suited for the EChO application.

The process of removing the CdZnTe-substrate, commonly used by the MCT detectors, has become standard. This allows the extension of the cut-on wavelength of these detectors to shorter ranges down to about 0.4{\mm}. Therefore, these detectors can be used for the full spectral coverage of the VIS- up to the IR3-channel. This approach provides advantages with respect to commonality and spare policy aspects. Alternatively, the VIS-channel could be equipped with a Hybrid Visible Silicon PIN detector array (HyViSi) with the same Read-Out Integrated Circuit (ROIC) and the same read-out scheme like the MCT detectors. However, beside the slightly higher quantum efficiency at 0.4{\mm}, no other significant advantages are known and therefore MCT detectors are preferred.

For the architecture of the ROIC the most commonly used input stage is the source follower. For low and medium detector currents, it offers the advantages of very low noise and low power dissipation. This type of ROIC is the standard for MCT detectors for the visual and near IR wavelength bands up to \ap5{\mm}. A disadvantage of the source follower circuit is the variation of detector bias voltage during an integration due to the increasing charge accumulation. During a reset, the bias voltage jumps back to the initial value. Nonetheless, this effect is negligible for detectors with low dark current and with relatively high bias voltages (in the 1 V-range and above). Therefore, the MCTs for cut-off wavelengths up to 5{\mm} and also the Si:As IBC/BIB detectors can use this type of ROIC.
 
All six detectors will be operated by a dedicated FEE. The FEE will be located close to the FPA and will provide all the required DC voltages, clocks, pulse patterns and control signals to the ROIC. It will also digitize the analog video output signals with 18 bit ADCs and provide the data stream via LVDS interface to the warm electronics. This approach will keep the sensitive analog lines short and therefore, reduce the susceptibility to electro-magnetic interferences. The FEE will be operated at cryogenic temperatures around 55~K. For the actual instrument design, the use of Teledyne's SIDECAR ASICs is intended \cite{Loose:2005}, which have a high technology readiness and space heritage. The SIDECAR ASIC provides high flexibility and can be configured for all read-out modes intended for this instrument.

%---------------------
\subsection{Flat-field calibration sources}\label{sec:fcs}
To characterize the pixel-to-pixel gain variations to a local spatial accuracy of 0.5\% (see Sect.~\ref{sec:sampling}), flat field calibration measurements should be performed frequently. Two dedicated on-board flat-field calibration sources (FCS) provide the required flat illumination of the detectors; one for the short wavelength channels up to 5{\mm} (SW-FCS) and one for the long wavelength channels (LW-FCS).

\begin{figure}[h]
\begin{center}
\includegraphics[width=0.4\textwidth]{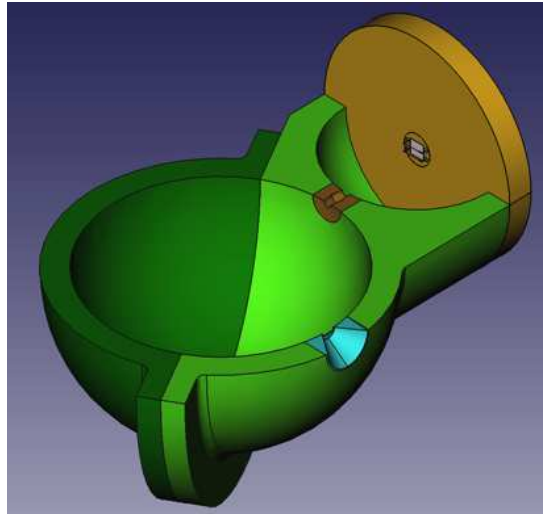} 
\end{center}
\caption{The FCS design shown in a cutaway view}
\label{fig:fcs}
\end{figure}

The requirements for both FCSs are very similar with the exception of the operating wavelength specification. Effort has been made to design two similar sources that would benefit from the same technology and operating modes. The concept is based on the sources used for the Mid InfraRed Instrument (MIRI) of the James Webb Space Telescope \cite{Glasse2006}. These sources use two identical tungsten filaments (to provide redundancy) mounted above a diffuser plate and integrated into a semi-sphere to provide an output intensity profile which is flat and homogeneous. The semi-sphere as well as the plate closing the semi-sphere are gold coated mirrors. To increase the spatial homogeneity of the source and to control the output flux of the system, the semi-sphere is attached to an integrating sphere with an intermediate aperture assembly. Fig.~\ref{fig:fcs} shows a mechanical drawing of the FCS design.
The filaments of the FCS are operated at 1600~K for the SW-FCS and at 700~K for the LW-FCS. This is not sufficient to cover the full wavelength range since the blackbody emission of a 1600~K source is too faint for the visible spectrum. Alternative light sources such as LEDs are studied for the VIS channel.

The light of the two sources is collimated by designated Cassegrain telescopes. The SW-PAS combines the collimated beams of the SW-FCS and the telescope (science beam) at the last PAS mirror. This mirror is back-illuminated by the FCS and acts as a central obscuring mask. In this way, the FCS beam builds ``a penumbra'' around the science beam. Since the FCS beam is injected before the dichroic mirrors, the calibration light is also dispersed. This provides a flat-field illumination with a representative spectrum compared to the science target (the same color and the same field, but not necessarily the same illumination). This shall enable that the flat-field measurement can probe color dependent gain effects and potentially provides a correction method for color dependent signal modulations caused by instrument-internal interferences.

The light of the SW-FCS will be blocked by the pupil-stop of the LW-PAS. Identically to its short wavelength counterpart, the LW-FCS is then injected at the last mirror of the LW-PAS.
%---------------------
\subsection{Cooling concept}\label{sec:cooling}
The instrument thermal architecture includes five thermally isolated environments: 
\begin{enumerate}[ (1)]
\item The main instrument optical bench (IOB) with the SW-PAS, SW-FCS, dichroic chain and the channels VIS, IR1, IR2, and IR3 (without detectors) at 55~K,
\item the MCT detectors of the channels VIS, IR1, IR2, IR3 at 55~K,
\item the cold optical bench including the LW-PAS, LW-FCS and the channels IR4 and IR5 (without detectors) at 15~K,
\item the Si:As detectors of the channels IR4 and IR5 at 7~K, and 
\item the FEE for all channels, at 55 -- 65~K.
\end{enumerate}
The thermal decoupling of the channel optics from its detectors and FEEs is primarily motivated by the high susceptibility of these devices to thermal variations. With their dedicated thermal environment, high precision thermal control can be enabled.

Each of the five thermal environments will have its dedicated cooling system. The spacecraft thermal control system provides passive cooling of the telescope down to 55~K using three V-grooves at different temperature levels (as indicated by \citet{Puig:2012dc}). These V-grooves are layers of thermal radiation insulation, placed between the warm part of the spacecraft and the cryogenic telescope. They are shaped in a V-form to shield the telescope most efficiently for different orientations of the spacecraft with respect to the sun. The IOB and its associated optical elements will be thermally coupled to the inner V-groove, providing sufficient cooling power for the dissipated heat of about 134~mW (including parasitic loads). An additional radiator area dedicated to the instrument will be used to cool the MCT detectors (24~mW) and FEEs (70~mW) passively. 

The cold optical bench at 15~K and the Si:As detectors at 7~K require active cooling. Different cooling concepts have been studied and further studies are required to adjust the cooling power with the available pre-cooling capacities from the spacecraft's V-grooves. The baselined cooler concept foresees to use a modified version of the 4~K Joule-Thomson cooler used for the Planck mission \cite{Bradshaw:1997}. The required pre-cooling at 15~K will be conducted by a vibration-free sorption cooler based on the technology development for the Darwin and IXO missions \cite{terBrake:2011ti}. Both coolers show sufficient cooling capacity, in accordance with the present state of the spacecraft thermal architecture.

%---------------------
\subsection{On-board electronics and data handling}
The instrument will be equipped with an instrument control unit and a separate control unit dedicated for the cryocoolers. The instrument control unit will be responsible for the power supply, the commanding, and the data link to the spacecraft and FEEs. Furthermore, this unit will drive the thermal control heaters and collect housekeeping information. The calibration sources will be driven by this control unit too. And finally, the instrument control unit will be responsible for the data handling.

According to a first estimation of telemetry rates, the instrument will produce an average raw data rate of about 38~GBit~/~day. The current allocation for spacecraft-downlink rates is around 5~GBit~/~day. Even with loss-less compression, the data rate is still too high for the  downlink budget and requires further reduction. Consequently, a dedicated data processing unit with sufficient computing power is required to compress the data. 

The data rate is higher with brighter targets as the detectors have to be reset more frequently. Depending on the source brightness, the detector read-out scheme consists of either correlated double sampling (CDS, for bright targets with integration times of 4~s or less) or non-destructive reads (NDR) by sampling-up-the-ramp (for fainter targets with up to 8 samples per ramp). The on-board data processing includes averaging of CDS data, or in the case of NDR data, fitting slopes to the samples. If required, the slopes of several integrations might be averaged to further reduce the time resolution and the data rate.

On-board data processing is mandatory to comply with the downlink rates. However, this data compression implies some loss of information, mainly by giving up a high time resolution. Even more critical is the effect of slope fitting: Modulations of the signal for individual pixels on timescales shorter than an integration (caused by, e.g., the pointing-jitter) lead to non-linear slopes. An automated fitting routine will be limited in correcting for such drifts and consequently, systematic errors are expected to be non negligible. With the requirement for photometric stability though, systematic errors have to be reduced to a very low level. Therefore, the definition of the on-board processing algorithms requires substantial development effort to guarantee that performance-critical information is maintained. 
 
%---------------------
\subsection{An alternative instrument concept -- passive cooling only}
Active cooling has significant implications on the mission cost and complexity. Preliminary cost estimates show that an actively cooled instrument for EChO is by 20\% more expensive compared to a concept with passive cooling only. However, one important factor to consider is the possible wavelength coverage and the science capability of a purely passively cooled instrument: If no active cooler is on board of the EChO spacecraft, the V-grooves are not used for pre-cooling the different stages of the cooler(s). Therefore, the temperature levels of the V-grooves are lowered and the telescope and the instrument reach temperatures around \ap45~K. With  the dedicated instrument radiator, the MCT detectors could then be cooled down to 38~K. This allows using long wavelength MCT detectors up to 10.5{\mm} \cite{McMurtry:2013ww}, which have still acceptable low dark currents ($< 200~\textrm{e}^-/\textrm{s}/\textrm{pixel}$).

Such an alternative design implies the reduction of wavelength coverage by removing channel IR5. The wavelength ranges of the remaining channels require some reallocation to cover the spectrum up to 10.5{\mm}. All detectors could be operated at the same temperature. The resulting thermal design is much simpler and the technological and programatic risks associated with the cryocooler are much reduced. 

On the other hand, the decreased wavelength coverage from 16 to \ap11{\mm} implies a reduction of scientific capabilities. The primary goal of observing at these long wavelengths is to measure thermal emission of cool objects, and especially to detect the 15{\mm} CO$_2$ feature in temperate super Earth planets. However, transmission spectroscopy at shorter wavelength is more sensitive to all molecules in the atmospheres of such objects. We have simulated observations of temperate super Earths of the ESA reference sample. In the emission spectrum EChO clearly detects the 15{\mm} CO$_2$ feature, but other species are not detected. We investigated the emission spectroscopy case in great detail, and conclude that there is only limited potential for discovering molecules at long wavelengths that lack features at shorter wavelengths. Thus, there is only marginal scientific merit to the 11--16{\mm} wavelength range, and such an alternative concept is indeed an attractive option. 

%===================================================================================================

\section{Simulated Performance Results}\label{sec:performance}
A first set of performance analyses was conducted as part of the assessment study. In particular, mission critical parameters such as the instrument transmission and photometric performance were derived. In the following we highlight a few results.

%---------------------

\subsection{Photon conversion efficiency}
\begin{figure}[h]
\begin{center}
\includegraphics[width=0.5\textwidth]{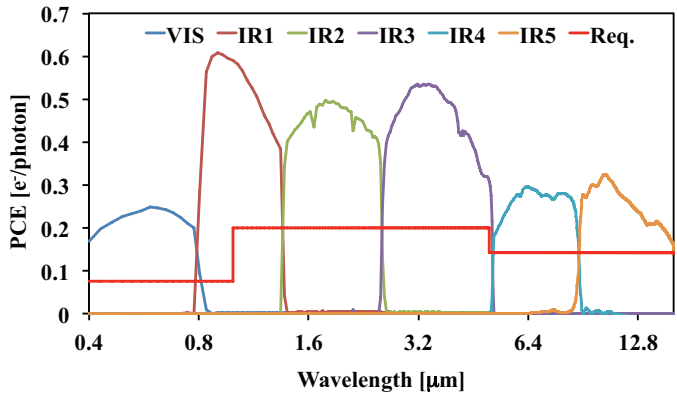} 
\end{center}
\caption{Photon conversion efficiency of the instrument compared with the mission requirement (red line).}
\label{fig:pce}
\end{figure}

The photon conversion efficiency (PCE) is the product of the instrument optical transmission and the quantum efficiency of the detectors. It is therefore the best measure for a telescope independent figure of merit. Fig.~\ref{fig:pce} shows the PCE for all channels and the comparison with the mission requirements.
The calculations include conservative figures of the reflectivity of the mirror surfaces, the transmission and reflectance of the dichroic mirrors, the efficiencies of the gratings and the quantum efficiencies of the focal plane arrays based on information from the supplying companies. The PCE shows for all wavelength sufficient margin with respect to the minimum requirement.

%---------------------
\subsection{Photometric model}
We developed an instrument dedicated photometric model \cite{vanBoekel:2012fy}. The radiometric model includes a set of stellar photosphere and planetary atmosphere models that are used to calculate flux levels and eclipse depths, as well as a model for the zodiacal background. 

\begin{figure}[h]
\begin{center}
\includegraphics[width=0.5\textwidth]{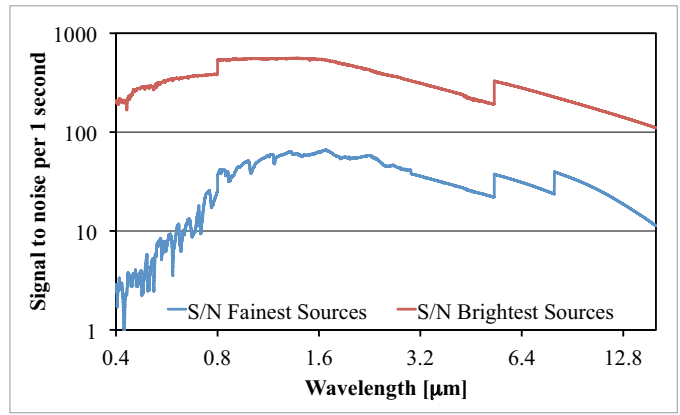} 
\end{center}
\caption{S/N for 1 second integration time for the brightest stars (K0V, $K_\textrm{mag}=4.0$, red line) and faintest stars (for $\lambda<3${\mm}: M5V, $K_\textrm{mag}=8.8$, for 3{\mm} $\le\lambda\le 8${\mm}: G0V, $K_\textrm{mag}=9.0$, and for $\lambda>8${\mm}: G0V, $K_\textrm{mag}=8.8$, blue line).}
\label{fig:sn}
\end{figure}

Based on the PCE, the estimated performance of the telescope, and the detector characteristics, we calculate the signal and the noise for each detector pixel. Integrating over all pixels of a spectral resolution element provides us with the resulting S/N. In particular, we derive the sensitivity performance for the bright and faint star limits (see Table~\ref{tab:key_figures_ECHO}) to verify compliance with the sensitivity requirements. Fig.~\ref{fig:sn} shows the S/N for an integration time of 1 second of the stellar spectrum (including the zodiacal background) for these two cases.
For all wavelengths, the S/N is dominated by the stellar photon noise with the exception for the faintest sources and for longer wavelengths where the zodiacal light becomes dominant. Nonetheless, the sensitivity performance is compliant to the requirement of photon noise limitation (of non instrumental photons, see next section). Consequently, the S/N can be scaled accordingly with the square root of the effective exposure time. The discontinuous transition of the S/N curve at 5{\mm} can be explained by the lower spectral resolving power of the channels IR4 and IR5. At 8{\mm}, the definition of the faint source limit is discontinuous (see Table~\ref{tab:key_figures_ECHO}), therefore the S/N plot shows a discontinuity as well.

\subsection{Photometric stability and noise budget}\label{sec:noisebudget}
As discussed in Sect.~\ref{sec:xfact}, we set up a Dynamic Performance Calculator to combine the various system noises and calculate their propagation. We use the introduced terminology of the $X$-factor to express the contribution of the different noise terms to the overall instrument performance with respect to photometric stability. Table~\ref{tab:x_plot} shows the resulting contribution to the photometric variability for different flux levels and wavelengths.
\begin{table}[h]
\caption{Photometric noise contributions expressed in $X$ for the bright and faint sensitivity limits at different wavelengths}
\label{tab:x_plot}
\begin{center}
\begin{tabular}{@{}l r r r r r r r r}
\toprule
Wavelength &	\multicolumn{2}{c}{0.4\mm}	&	\multicolumn{2}{c}{1\mm}	&	\multicolumn{2}{c}{11\mm}	&\multicolumn{2}{c}{16\mm}		\\
		&	bright	&	faint	&	bright	&	faint	&	bright	&	faint	&	bright	&	faint	\\
 \colrule

Telescope Background	&	0.00\%	&	0.00\%	&	0.00\%	&	0.00\%	&	0.01\%	&	0.07\%	&	6.44\%	&	31.34\%	\\
Read-Out-Noise		&	0.14\%	&	0.41\%	&	0.01\%	&	0.00\%	&	0.43\%	&	0.01\%	&	0.78\%	&	0.01\%	\\
Dark Current			&	0.00\%	&	5.92\%	&	0.00\%	&	0.04\%	&	0.02\%	&	0.23\%	&	0.04\%	&	0.19\%	\\
Detector Gain Variation	&	0.01\%	&	0.00\%	&	0.15\%	&	0.00\%	&	0.04\%	&	0.00\%	&	0.02\%	&	0.00\%	\\
Slit-losses			&	0.01\%	&	0.00\%	&	0.11\%	&	0.00\%	&	3.30\%	&	0.32\%	&	1.85\%	&	0.38\%	\\
On-Board Processing	&	1.00\%	&	1.00\%	&	1.00\%	&	1.00\%	&	1.00\%	&	1.00\%	&	1.00\%	&	1.00\%	\\
Uncorrected Detector drifts&	3.00\%	&	3.00\%	&	3.00\%	&	3.00\%	&	3.00\%	&	3.00\%	&	3.00\%	&	3.00\%	\\
Astronomical noise		&	3.00\%	&	3.00\%	&	3.00\%	&	3.00\%	&	3.00\%	&	3.00\%	&	3.00\%	&	3.00\%	\\
Margin (20\%)		&	1.43\%&	2.66\%	&1.45\%&	1.41\%	&2.16\%	&1.53\%&	3.23\%&	7.78\%\\
 \colrule
Total	 ($X$)			&	8.60\%&	15.99\%	&8.72\%	&8.45\%	&12.97\%	&9.16\%	&19.35\%&	46.71\%\\
\botrule
\end{tabular}
\end{center}
\end{table}

The telescope background was estimated by assuming a mirror temperature of 55~K and an emissivity of $\epsilon=97\%$. The detector dark current and read noise were estimated conservatively using preexisting information from the supplying companies. The contributions from the pointing-jitter in combination with detector gain variations and slit-losses were addressed in Sect.~\ref{sec:trade-off} and explicitly modeled with respect to noise type A and B contributions. 

In Table~\ref{tab:x_plot}, a few more rows are shown with uniform $X$-values: For the information loss due to on-board processing, a generic systematic error of $X=1\%$ is assumed. This value needs further evaluation to better quantify the effective systematic errors introduced. Further, we budget a generic $X$=3\% contribution for any uncorrectable detector effect that causes drifts in the signal responses. This value needs also further refinement using dedicated laboratory measurements.

It seems inappropriate to define an astronomical noise term within an instrumental noise budget. However, this is a result from the formulation of the photometric stability requirement: To achieve the fundamentally limited signal-to-noise, only targets with low stellar variability should be observed for which the noise contribution is lower than $X$=3\% (after de-trending the time series using stellar variability models). Consequently, we budget $X$=3\% as an upper limit for the correction of stellar variability.

\begin{figure}[h]
\begin{center}
\includegraphics[width=0.5\textwidth]{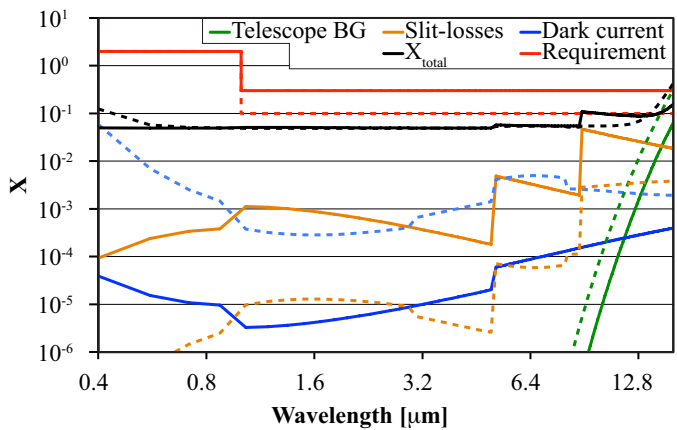} 
\end{center}
\caption{Photometric stability performance expressed as $X$-factor for the brightest (solid line) and the faintest sources (dashed line). The individual contributions from the telescope thermal background, slit-losses and dark current are shown. System noise components with constant $X$-values are not shown (see Table~\ref{tab:x_plot}). The required values for $X$ are plotted in solid red, the goal values are plotted in dashed red.}
\label{fig:X}
\end{figure}

All these noise terms added together including an additional margin of 20\% are compared with the required upper limit for $X$. Fig.~\ref{fig:X} shows the result graphically for the full wavelength range.
	
In general, sufficient margin to the required level can be demonstrated with the exception of the longest wavelengths: At wavelengths larger than \ap13{\mm}, the telescope background dominates the noise budget and exceeds the allocation for the faintest targets.

With the presented selection of detectors, the resulting dark current plays a minor role for the final noise floor. Around 9{\mm} the effect of slit-losses dominates the noise floor for the brightest objects, leading to a total noise which exceeds slightly the goal value of $X=10\%$. However, to reduce this noise component, a larger slit width would be needed. This would increase the contribution from the telescope thermal background, exceeding the allocated budget for even shorter wavelengths. Consequently, the presented slit width of 13~arcsec represents a good compromise between the performance at 9{\mm} and the performance at wavelengths greater than 13{\mm}.

\section{Conclusions and Outlook}\label{sec:outlook}
We have demonstrated how the key mission requirements of the EChO mission allow discriminating different spectrometer concepts of which the multichannel dispersive and all-reflective spectrometer appears optimal. To achieve the high photometric stability required for this mission, the design of the spectrometer is driven by various fundamental considerations. The optical design is fully optimized to achieve optimally stable performance, including the definition of the FoV, the spatial sampling, and the beam broadening elements. The selection of the detector technologies is based on the constraints of the acceptable dark currents and stability performance. The thermal architecture allows decoupling of temperature sensitive elements from the rest of the instrument, providing a better environment to achieve high thermal stability. With the active cooling required to cover the longest wavelengths up to 16{\mm}, the mission becomes more complex and expensive. For this reason we presented in addition an alternative approach based on passive cooling only but with a reduced wavelength range up to \ap11{\mm}. 

The presented study illustrates the technical feasibility of a spectrometer instrument for the EChO observatory. Performance estimations demonstrate compliance of the system with all mission requirements and in particular with the challenging photometric stability criterion. This result will support the overall feasibility assessment of the EChO mission to be concluded in 2013. After the selection of the instrument consortium, the mission is now subject to a selection process in competition with three other candidates for the M3 mission of the  Cosmic Vision 2015-2025 program. This selection process is expected to conclude early 2014, followed by a detailed design study in Phase B.

\section*{Acknowledgments}
We would like to thank our industrial partners for their support of this study: Astrium GmbH, Kayser-Threde GmbH, AIM, brusag Sensorik \& messtechnische Entwicklungen AG, Karlsruhe Institute for Technology, Holota Optics, Fraunhofer IOF, ZEISS Oberkochen, RUAG Space, Teledyne Technologies Inc., DRS Technologies Inc. We like to thank M. ter Brake (University Twente) and T. Bradshaw (Rutherford Appleton Laboratory) for their  support concerning the active cooler concept. Further, we would like to thank the ESA EChO study team for their support and guidance during this study. 

This work has been supported by the German Space Agency (DLR) with funds provided by the Federal Ministry of Economics and Technology (BMWi) under grant number 50 OO 1202 and by the Swiss Prodex Program under grant number 4000105381.

N.~Madhusudhan acknowledges support from the Yale Center for Astronomy and Astrophysics (YCAA) at Yale University through the YCAA prize postdoctoral fellowship.

%\bibliographystyle{ws-jai}
%\bibliography{refs}   

\end{document}